\documentclass[10pt,aps,amsmath,amssymb, prd,nofootinbib, twocolumn,preprintnumbers,superscriptaddress]{revtex4-2}

\usepackage{graphicx}
\usepackage{amsmath}
\usepackage[caption=false]{subfig}
\usepackage{siunitx}
\usepackage{placeins}
\usepackage{color}
\usepackage{standalone}
\usepackage{dcolumn}
\usepackage{tensor}
\usepackage{bm}
\usepackage{microtype}
\usepackage{etoolbox}
\usepackage{amssymb}
\usepackage{mathrsfs}
\usepackage{accents}
\usepackage[normalem]{ulem}
\usepackage{soul} 
\usepackage[dvipsnames]{xcolor}
\usepackage[colorlinks,urlcolor=Blue,citecolor=Blue,linkcolor=Blue,pdfusetitle]{hyperref}
\usepackage[all]{hypcap}
\usepackage[inline]{enumitem}
\usepackage[utf8]{inputenc}
\usepackage{xspace}
\usepackage[printonlyused, nolist]{acronym}
\usepackage{lineno}
\usepackage{bbding}
\usepackage{pifont}
\usepackage{wasysym}
\usepackage{xspace}

\usepackage{tablefootnote}
\usepackage{booktabs}
\usepackage{longtable}
\usepackage{color, colortbl}
\definecolor{Color1}{rgb}{0.9453,0.9453,0.9687}
\definecolor{Color2}{rgb}{0.8476,0.8945,0.9375}
\definecolor{Color3}{rgb}{0.7461,0.8164,0.8945}
\definecolor{Color4}{rgb}{0.6484,0.7383,0.8554}
\definecolor{Color5}{rgb}{0.5,0.6484,0.8047}

\newcommand{\xmark}{\ding{55}}
\newcommand{\cmark}{\ding{51}}

\usepackage{orcidlink}

\begin{document}
\title{Validating Prior-informed Fisher-matrix Analyses against GWTC Data}

\author{Ulyana Dupletsa\,\orcidlink{0000-0003-2766-247X}}
\email{ulyana.dupletsa@gssi.it}
\affiliation{Gran Sasso Science Institute (GSSI), I-67100 L'Aquila, Italy}
\affiliation{INFN, Laboratori Nazionali del Gran Sasso, I-67100 Assergi, Italy}
\author{Jan Harms\,\orcidlink{0000-0002-7332-9806}}
\affiliation{Gran Sasso Science Institute (GSSI), I-67100 L'Aquila, Italy}
\affiliation{INFN, Laboratori Nazionali del Gran Sasso, I-67100 Assergi, Italy}
\author{Ken K.~Y.~Ng\,\orcidlink{0000-0003-3896-2259}}
\affiliation{William H. Miller III Department of Physics and Astronomy, Johns Hopkins University, Baltimore, Maryland 21218, USA}
\author{Jacopo Tissino\,\orcidlink{0000-0003-2483-6710}}
\affiliation{Gran Sasso Science Institute (GSSI), I-67100 L'Aquila, Italy}
\affiliation{INFN, Laboratori Nazionali del Gran Sasso, I-67100 Assergi, Italy}
\author{Filippo Santoliquido\,\orcidlink{0000-0003-3752-1400}}
\affiliation{Gran Sasso Science Institute (GSSI), I-67100 L'Aquila, Italy}
\affiliation{INFN, Laboratori Nazionali del Gran Sasso, I-67100 Assergi, Italy}
\author{Andrea Cozzumbo\,\orcidlink{0009-0004-2772-2692}}
\affiliation{Gran Sasso Science Institute (GSSI), I-67100 L'Aquila, Italy}
\affiliation{INFN, Laboratori Nazionali del Gran Sasso, I-67100 Assergi, Italy}

\begin{abstract}
Fisher-matrix methods are widely used to predict how accurately parameters can be estimated. Being computationally efficient, this approach is prompted by the large number of signals simulated in forecast studies for future gravitational-wave (GW) detectors, for which adequate analysis tools and computational resources are still unavailable to the scientific community. However, approximating the full likelihood function with a Gaussian may lead to inaccuracies, which we investigate in this work. To assess the accuracy of the Fisher approximation, we compare the results of the Fisher code \texttt{GWFish} against real data from the Gravitational Wave Transient Catalogs (GWTCs) provided by the Virgo/LIGO Bayesian analyses. Additionally, we present a sampling algorithm to include priors in \texttt{GWFish}, not only to ensure a fair comparison between \texttt{GWFish} results and the Virgo/LIGO posteriors but also to investigate the role of prior information and to assess the need to include it in standard Fisher analyses. We find that the impact of priors depends mostly on the level of signal-dependent degeneracy of the waveform parameterization, and priors are generally more important when the level of degeneracy is high. Our findings imply that Fisher-matrix methods are a valid tool for ET science-case studies.
\end{abstract}

\maketitle
\section{Introduction}
Next-generation gravitational wave (GW) observatories, such as the Einstein Telescope (ET) \citep{Hild:2008ng, Punturo_2010, Hild:2010id} and the Cosmic Explorer (CE) \citep{Reitze:2019iox, Evans:2021gyd}, are attracting numerous studies to assess their performance and scientific potential. The wealth of data these instruments are expected to provide \citep{Maggiore:2019uih}, compared to past and ongoing observations \citep{LIGOScientific:2016aoc, LIGOScientific:2018mvr, LIGOScientific:2020ibl, KAGRA:2021vkt}, poses a computational challenge currently addressed by Fisher-matrix techniques \citep{Branchesi:2023mws}. However, the reliability of the Fisher approximation is still an open question. 
In what follows, we aim at investigating the potentialities and limitations of using the Fisher matrix by comparing its results to the standard analysis in the GW field performed in the LIGO-Virgo-KAGRA (LVK) Collaboration.

GW data analysis in the LVK Collaboration relies on Bayesian parameter estimation \citep{Ashton:2018jfp, Romero-Shaw:2020owr, Ashton:2021anp}. Instead, Fisher-matrix analysis \citep{Cutler:1994ys,Vallisneri:2007ev} has become a customary approach to simulate the precision of parameter estimation for next-generation detectors. 
In recent years, various Fisher-matrix codes have been developed, in particular \texttt{GWFish} \citep{Dupletsa:2022scg}, \texttt{GWBench} \citep{Borhanian:2020ypi}, \texttt{GWFast} \citep{Iacovelli:2022bbs, Iacovelli:2022mbg}, and \texttt{TiDoFM} \citep{Li:2021mbo, Chan:2018csa}. The main differences between the codes lie in the methods used for differentiation and matrix inversion, which are crucial steps in the analysis. Nevertheless, the codes have undergone several successful cross-checks and output comparable results \citep{Branchesi:2023mws}.

Even though Fisher-matrix analysis is an approximate 
technique based on Gaussian likelihood approximation expected to hold just for high signal-to-noise (SNR) events \citep{Vallisneri:2007ev}, its strengths lie in being computationally fast (seconds of computational time against days for one event) and thus effective in large-scale simulations.

Moreover, the standard Fisher analysis assumes the priors on the parameters to be uniform. In the high-SNR limit, the likelihood is expected to be a narrowly peaked function around the maximum. Therefore, no prior information is needed.
In practice, this might happen in just a few cases. Correlations among parameters rather than low SNR can limit the accuracy of parameter errors obtained under the Gaussian likelihood approximation without priors. This is because correlations can create wide likelihoods. 
Therefore, we look into the effect of including prior information so that posterior samples do not overestimate errors by orders of magnitude \citep{Rodriguez:2013mla} potentially going outside the physical range of parameters. 

The influence of including priors into standard Fisher analysis was introduced in~\citep{Vallisneri:2007ev}, and an implementation via a rejection sampling algorithm has been proposed in~\citep{Iacovelli:2022bbs}. 
Some hints to the importance of priors implementation are already present in \citep{Branchesi:2023mws}, mostly related to the inclination angle prior in the context of multi-messenger studies with binary neutron star systems. Furthermore, including priors is being exploited to ensure reliable distance errors, e.g., with a rejection sampling method in the work of \citep{Santoliquido:2024oqs} for machine learning classification and with the method proposed in our work in \citep{Cozzumbo:2024vxw} to reconstruct the expansion history of the universe in the bright sirens framework. However, we recognize that if there are multimodalities, they are not generally suppressed by including priors, as we show in this work. Multimodalities, which could arise from cyclic parameters, pose a delicate issue in the analysis.

The scope of this paper is twofold. First, we describe a computationally efficient method to incorporate priors into our Fisher-matrix code, \texttt{GWFish} \citep{Dupletsa:2022scg}. Compared to methods where parameter-estimation errors are calculated directly from Fisher matrices, sampling from a Gaussian likelihood with priors only takes about twice the computational time\footnote{Assuming we are drawing an order of magnitude of $10^4$ samples. See also discussion in~\ref{sec:sampling}.}. Second, we want to stress the role priors play both in the approximate Fisher analysis and the standard Bayesian approach. This is done by consistently comparing our prior-informed Fisher results to the full Bayesian results for all of the available GW data from the public Gravitational Wave Transient Catalogs (GWTCs) \citep{LIGOScientific:2018mvr, LIGOScientific:2020ibl, KAGRA:2021vkt}. 

We structure our work as follows. In Sec.~\ref{sec:analysis}, we give a brief but comprehensive overview of the analysis technique adopted in our work. In Sec.~\ref{sec:gwtc_cfr}, we test our analysis framework against GWTCs data. In Section~\ref{sec:discussion}, we provide a detailed discussion of our study's findings and highlight the potential and limitations of data-analysis techniques in the GW field. We pay particular attention to the role that assumptions on priors play.

\section{Analysis}
\label{sec:analysis}

\subsection{Parameter Estimation}
Properties of astrophysical sources from observed GW signals are inferred through Bayesian analysis \citep{Thrane_2019, Christensen:2022bxb}. This hinges on the Bayes theorem, where the posterior distribution of a given parameter is obtained by combining the prior information on the parameters  $\pi(\vec{\theta})$ with the likelihood of observing the given data  ${\mathcal{L}}(d|\vec{\theta})$:
\begin{equation}
    \label{eq:bayes}
    p(\vec{\theta}|d) \propto \pi(\vec{\theta}){\mathcal{L}}(d|\vec{\theta}),
\end{equation}
where $\vec{\theta}$ denotes the parameters entering the modelling of the GW signal. The data $d$ is assumed to consist of both noise $n$ and the true signal $h(\vec{\theta}_0)$, which depends on the unknown true parameters $\vec{\theta_0}$:
\begin{equation}
    d = n + h(\vec{\theta}_0).
\end{equation}
$h$ represents the waveform approximant chosen to model the data. The noise is assumed to be Gaussian and stationary.
Indeed the likelihood ${\mathcal{L}}(d|\vec{\theta})$ is defined as the probability of a noise realization, meaning that it quantifies the probability that the observed data $d$ could be a specific realization of the noise in the detector:
\begin{equation}
    \label{eq:lkh}
    {\mathcal{L}}(d|\vec{\theta}) \propto \exp\left[-\frac{1}{2}\langle d - h(\vec{\theta})| d - h(\vec{\theta}) \rangle \right].
\end{equation}

The inner product $\langle \cdot|\cdot\rangle$ measures the overlap between two signals given the noise characteristics of the detector:
\begin{equation}
    \label{eq:inner_prod}
    \langle a| b \rangle \equiv 4\operatorname{Re}\int_{f_{\rm min}}^{f_{\rm max}} \frac{\tilde{a}(f)\tilde{b}^*(f)}{S_n(f)}df,
\end{equation}
with $S_n(f)$ the power spectral density of the detector noise. $\tilde{a}(f)$ and $\tilde{b}^*(f)$ are the Fourier transforms of the time-domain signals $a(t)$ and $b(t)$, and $f_{\rm min}$, $f_{\rm max}$ the frequency range determined by the sensitivity band of the detector. 

The standard Bayesian analysis estimates the signal parameters, employing Markov Chain Monte Carlo (MCMC) or Nested Sampling algorithms to explore the parameter space and efficiently sample the posterior.

In the Fisher matrix approximation, the likelihood is assumed to be a multidimensional Gaussian. This technique offers a quick and analytic way to estimate the covariance matrix of parameters without offering a straightforward way to consider prior information. 

In the following, we will briefly review the Fisher matrix formalism \citep{Vallisneri:2007ev} and then show how to enhance a standard Fisher matrix analysis in post-processing by incorporating physically motivated priors into the posterior samples. To this scope, we will use the \texttt{GWFish} \citep{Dupletsa:2022scg} Fisher matrix software and implement a sampling procedure to obtain posteriors, which can include both non-Gaussian and non-uniform priors.

\subsection{Fisher Matrix Formalism}
The Fisher matrix is defined as:
\begin{equation}
    \label{eq:fisher}
    F_{ij} = \Big\langle \frac{\partial h}{\partial \theta_i}\bigg| \frac{\partial h}{\partial \theta_j}\Big\rangle\bigg|_{\theta_{\rm inj}},
\end{equation}
where $h$ is the signal model, $\vec{\theta}_{\rm inj}$ are the injected parameters used to model the waveform signal $h(\vec{\theta}_{\rm inj})$, and the inner product is defined in Eq.~\eqref{eq:inner_prod}. In Fisher matrix analysis context the injected parameters $\vec{\theta}_{\rm inj}$ are the true parameters.

The inverse of the Fisher matrix represents the covariance matrix $\Sigma_{ij}$ of the parameters:
\begin{equation}
    \label{eq:cov_mat}
    \Sigma_{ij} = \left[ F^{-1}\right]_{ij}.
\end{equation}

The covariance matrix estimates the uncertainties and correlations in the parameters. The square root of the diagonal elements of $\Sigma_{ij}$ provides the $1\sigma$ uncertainties on the parameters, whereas the off-diagonal terms describe how different parameters are correlated.

\subsection{Integrating Priors into Fisher Matrix Analysis}
\label{sec:sampling}
The Fisher matrix analysis does not include information about the parameters' physical range, not to mention more complicated prior distributions. In the following, we outline the procedure to get full approximated posteriors, starting from the Fisher matrix. In a few words, this consists of drawing likelihood samples and weighting them by their prior probability. 
\begin{itemize}
    \item The Fisher matrix directly provides the multivariate Gaussian likelihood function:
    \begin{equation}
        \label{eq:fisher_lkh}
        {\mathcal{L}} \propto \exp\left[-\frac{1}{2} \left(\vec{\theta} - \vec{\theta}_{\rm inj}\right)^{\rm T} \Sigma^{-1} \left(\vec{\theta} - \vec{\theta}_{\rm inj}\right) \right],
    \end{equation}
    where $\Sigma$ is the inverse of the Fisher matrix and is directly calculated with \texttt{GWFish} as in Eq.~\ref{eq:fisher}. $\vec{\theta}_{\rm inj}$ are the injected (true) parameters. In our Fisher analysis, we work in the zero-noise approximation, therefore the likelihood in Eq.~\ref{eq:fisher_lkh} is centered on the injected values.
    \item Some parameters (especially angles and spins), even with high SNR, may still be poorly measured and show a largely-spread likelihood. Consequently, when drawing likelihood samples, most samples lie outside the physical range of parameters. Therefore, we sample directly from the truncated form of the Gaussian likelihood to prevent drawing a large number of samples with null prior, which would have to be discarded. In particular, we rely on the algorithm developed in \cite{2016arXiv160304166B}\footnote{We slightly modified the \texttt{python} implementation of \cite{2016arXiv160304166B} publicly available on \href{https://github.com/brunzema/truncated-mvn-sampler?tab=readme-ov-file}{\texttt{github}}}. This effectively accounts for the parameters' boundaries, ensuring the samples are in their physical range. When sampling from the truncated Gaussian we are not inverting again the covariance matrix. The algorithm, instead, uses the Cholesky decomposition. 
    \item For each sample realization drawn from the likelihood, we evaluate the prior probability (see \cite{Callister:2021gxf} and Appendix~\ref{app:choosing_priors} for a more detailed discussion).
    \item In the final phase of the analysis, we construct the posterior distribution. We start by evaluating the prior probability of each likelihood realization. We use \texttt{numpy}'s \texttt{random.choice} \citep{2020Natur.585..357H} function to select the prior re-weighted samples. The re-weighting process favours samples with higher prior probabilities.
\end{itemize}

The outlined approach combines the information from the Fisher analysis (via the likelihood) as from Eq.~\ref{eq:fisher_lkh} with the additional information provided by the priors following Eq.~\ref{eq:bayes}. Computationally speaking, including priors takes an overall time which is less than twice the one of the standard \texttt{GWFish} analysis to draw $10^4$ samples. This assures the efficiency of the prior-informed Fisher matrix method presented here. The code used to run the analyses in this work is publicly available on GitHub \href{https://github.com/u-dupletsa/GWFish-meets-Priors}{https://github.com/u-dupletsa/GWFish-meets-Priors}.

\section{Comparison with GWTC data}
\label{sec:gwtc_cfr}

We conducted an extensive comparison with real data to validate the need to include priors in standard Fisher analysis. To this scope, we recreated the same conditions used for the LVK data analysis and produced our Fisher matrix results.

\subsection{O1 + O2 + O3a+ O3b}
We used the updated version (\textit{v2}) of the LVK publicly available GW event database present on \texttt{Zenodo} \citep{zenodo} \footnote{\href{https://zenodo.org/records/6513631}{GWTC-2.1} for O1 + O2 + O3a and \href{https://zenodo.org/records/8177023}{GWTC-3} for O3b \label{footnote:lvk_zenodo_link}} including all the binary black hole (BBH) mergers from the first 3 observing runs \citep{LIGOScientific:2018mvr, LIGOScientific:2020ibl, KAGRA:2021vkt}. Specifically, we used the cosmologically reweighted posterior samples labelled as \textit{mixed\_cosmo}. In Tab.~\ref{tab:events_numbers} we report the comparison of the numbers between all the observed events and those we used in our analysis. In total, we have analyzed 78 events. For a comprehensive list of all the publicly available GW detections and why some of them were not included in our analysis, see Tab.~\ref{tab:events}.

\begin{table}[ht!]
  \caption[]{List of detected events during the first 3 observing runs by the LVK Collaboration, compared to those used in our analysis (see Tab.~\ref{tab:events} in Appendix~\ref{app:events} for further details).}
  \label{tab:events_numbers}
  \centering
  \begin{tabular}{l | c c c c | r}
  \toprule
    &\textbf{O1} &\textbf{O2} &\textbf{O3a} &\textbf{O3b} &\textbf{total} \\
    \midrule
    observed            &3  &7   &44   &36  &90\\
    included in our analysis    &3  &6   &36   &33  &78\\
    \bottomrule
  \end{tabular}
\end{table}

\subsection{Fisher analysis}
We analysed a total of 78 GW events (see Tab.~\ref{tab:events_numbers} and Tab.~\ref{tab:events}). The Fisher analysis of each event was carried out under the same conditions the full Bayesian analysis was obtained. This means:
\begin{itemize}
    \item Each event has its own detector network with specific sensitivity curves at the moment of detection.
    \item The waveform approximant is the same and is given by \texttt{IMRPhenomXPHM} \citep{Pratten:2020ceb}. This is one of the waveform approximants used in the LVK collaboration for BBH events, readily provided by \texttt{LALSimulation} \citep{Husa:2015iqa}.
    \item For each event, we take as injected values 30 different realizations\footnote{The number of realizations was chosen to be computationally feasible and provide enough statistics.} randomly sampled from the posterior distributions of the LVK analysis. This choice is motivated by the widely spread and often multi-modal posteriors LVK provides. Averaging over different realizations provides a more robust comparison.
    \item All the 15 parameters are included in the Fisher analysis: [${\mathcal{M}}_c$, $q$, $d_L$, $\theta_{JN}$, \texttt{DEC}, \texttt{RA}, $\phi$, $\Psi$, $t_c$, $a_1$, $a_2$, $\texttt{tilt}_1$, $\texttt{tilt}_2$, $\texttt{phi}_{12}$, $\texttt{phi}_{JL}$]. 
    The parameters represent the masses, in the form of chirp mass ${\mathcal{M}}_c$ (in detector frame) and mass ratio $q$, luminosity distance $d_L$, the inclination angle of the source with respect to the line of sight of the observer $\theta_{JN}$, parameters describing the position of the source in the sky (\texttt{DEC} and \texttt{RA}), the phase $\phi$ and the time of arrival of the signal $t_c$, its polarization $\Psi$, and the six parameters ($a_1$, $a_2$, $\texttt{tilt}_1$, $\texttt{tilt}_2$, $\texttt{phi}_{12}$, $\texttt{phi}_{JL}$) describing the spins of the two compact objects.
    See Tab.~\ref{tab:params_description} for a detailed description. 
\end{itemize}

\subsection{Full posterior reconstruction}
Starting from the Fisher results, we follow the method outlined in Sec.~\ref{sec:analysis} to obtain the posterior samples. The Fisher posterior samples, with and without the addition of priors, are then compared to the ones obtained with the full Bayesian analysis. Therefore, we have adopted the same priors as the full Bayesian procedure to ensure consistent comparison. The specific priors are reported in Tab.~\ref{tab:priors}. Since we have 30 different analyses for each event, the result of our comparison is a distribution.
\begin{table}[ht!]
  \caption[]{List of analytic priors for each parameter entering the description of a GW event. For a more detailed description of the parameters refer to Tab.~\ref{tab:params_description}.}
  \label{tab:priors}
  \begin{tabular}{l | c c c}
  \toprule
    \textbf{parameter} &\textbf{units} &\textbf{prior} &\textbf{prior range} \\
    \midrule
    ${\mathcal{M}}_c$ &$M_{\odot}$ &$\mathcal{M}_c$\footnote{\label{note0} Together with the prior on $q$, this ensures that the joint prior is uniform in the component masses, $m_1$ and $m_2$. See Appendix~\ref{app:choosing_priors} for further details.} &$[\text{min}, \text{max}]$\footnote{Each event has a specific prior range based on preliminary analysis. The prior interval can be more or less constrained in a case-dependent fashion.}\\
    $q$ &- &$q^{-6/5}(1+q)^{2/5}$\footref{note0} &$[0.05, 1.0]$\\
    $d_L$ &[Mpc] &Uniform in source-frame\footnote{See Appendix~\ref{app:choosing_priors} for further details.} &$[10, 10000]$\\
    $\theta_{JN}$ &[rad] &Sine &$[0, \pi]$\\
    \texttt{DEC} &[rad] &Cosine &$[-\frac{\pi}{2}, +\frac{\pi}{2}]$\\
    \texttt{RA} &[rad] &Uniform &$[0, 2\pi]$\footnote{\label{note1}The prior on these parameters has periodic boundary conditions, which cannot be implemented in the context of Fisher analysis.}\\
    $\phi$ &[rad] &Uniform &$[0, 2\pi]$\footref{note1}\\
    $\Psi$ &[rad] &Uniform &$[0, \pi]$\footref{note1}\\
    $t_c$ &[s] &Uniform &$[t_c - 0.1, t_c + 0.1]$\footnote{The prior is roughly plus and minus $0.1$ second around the estimated time of arrival, which is the injected value in case of Fisher matrix analysis. In the analysis, though, we have taken exactly the same range as the LVK analysis. This range is reported in the LVK database.}\\
    $a_1$ &- &Uniform &$[0, 0.99]$\\
    $a_2$ &- &Uniform &$[0, 0.99]$\\
    \texttt{tilt}$_1$ &[rad] &Sine &$[0, \pi]$\\
    \texttt{tilt}$_2$ &[rad] &Sine &$[0, \pi]$\\
    \texttt{phi}$_{12}$ &[rad] &Uniform &$[0, 2\pi]$\footref{note1}\\
    \texttt{phi}$_{JL}$ &[rad] &Uniform &$[0, 2\pi]$\footref{note1}\\
    \bottomrule
  \end{tabular}
\end{table}

\begin{figure*}
    \centering
    \includegraphics[scale=0.45]{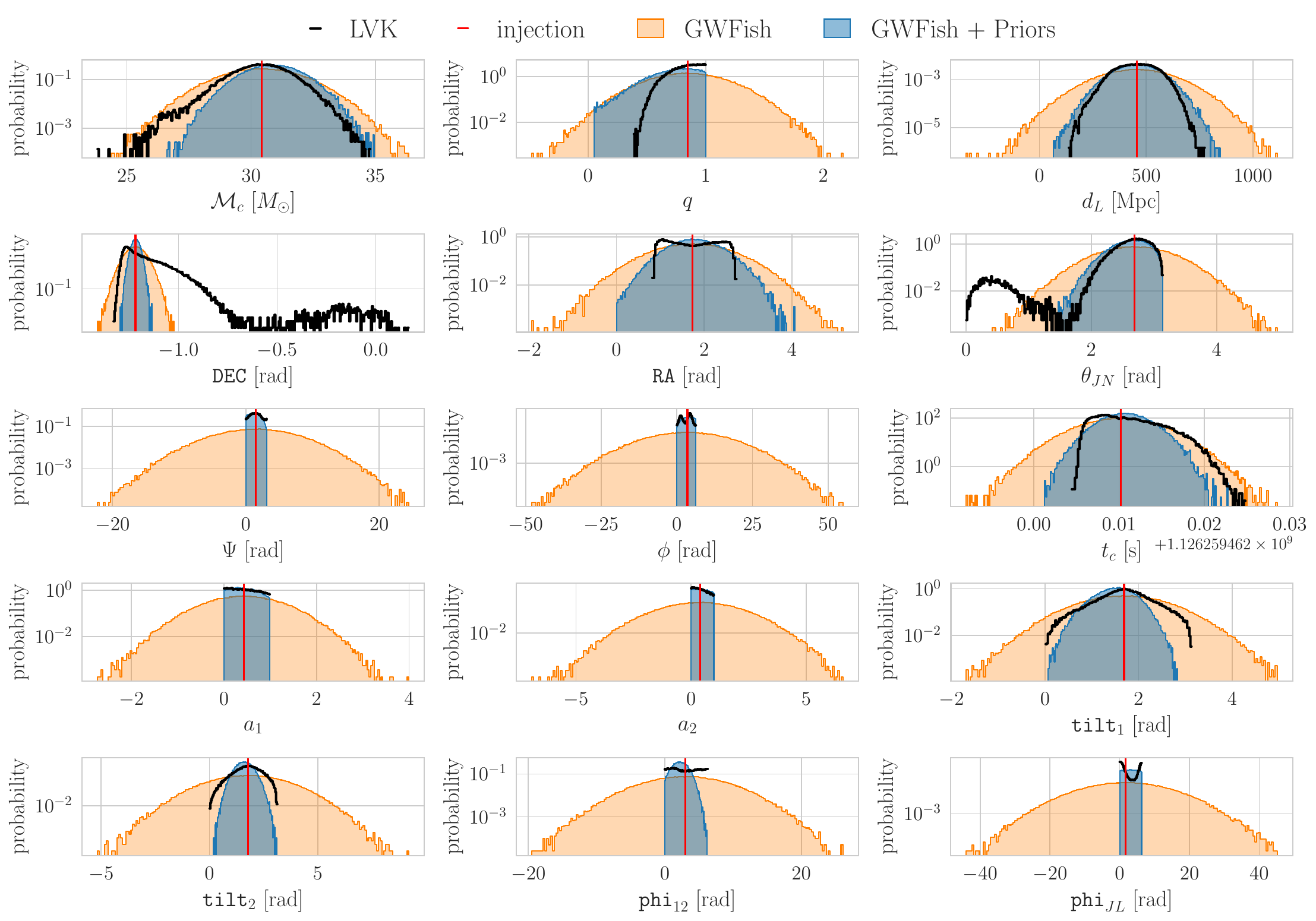}
    \caption{\texttt{GW150914} posterior distribution of all the 15 parameters. Full Bayesian parameter estimation results (LVK) are shown in \textit{black}, while \texttt{GWFish} results without (with) priors are shown in \textit{orange} (\textit{blue}). The vertical \textit{red} line is the median of the LVK posterior distribution, also used as injection value for Fisher analysis (see Eq.~\ref{eq:fisher_lkh})}
    \label{fig:posteriors_gw150914}
\end{figure*}

In Fig.~\ref{fig:posteriors_gw150914}, we show an example event, the first ever detected GW event, \texttt{GW150914}. We plot for each of the 15 parameters 3 sample sets: in \textit{black} the posterior samples provided by LVK, in \textit{red} the median value of the LVK distribution, which, in this example plot, we used as the injection for \texttt{GWFish}. We superimpose in \textit{orange} the samples obtained from the Fisher analysis alone (likelihood samples) as in Eq.~\ref{eq:fisher_lkh} and, in \textit{blue}, the samples we get after the inclusion of priors following Eq.~\ref{eq:bayes}. It is worth noting how including the prior information for some parameters, especially spins and angles, may play a crucial role and restrict the uncertainty estimates by a large amount (see Sec.~\ref{subsec:multi_modality} for a detailed discussion). For the parameter descriptions, refer to Tab.~\ref{tab:params_description}.

\begin{figure*}[ht!]
    \centering
    \includegraphics[scale=0.7]{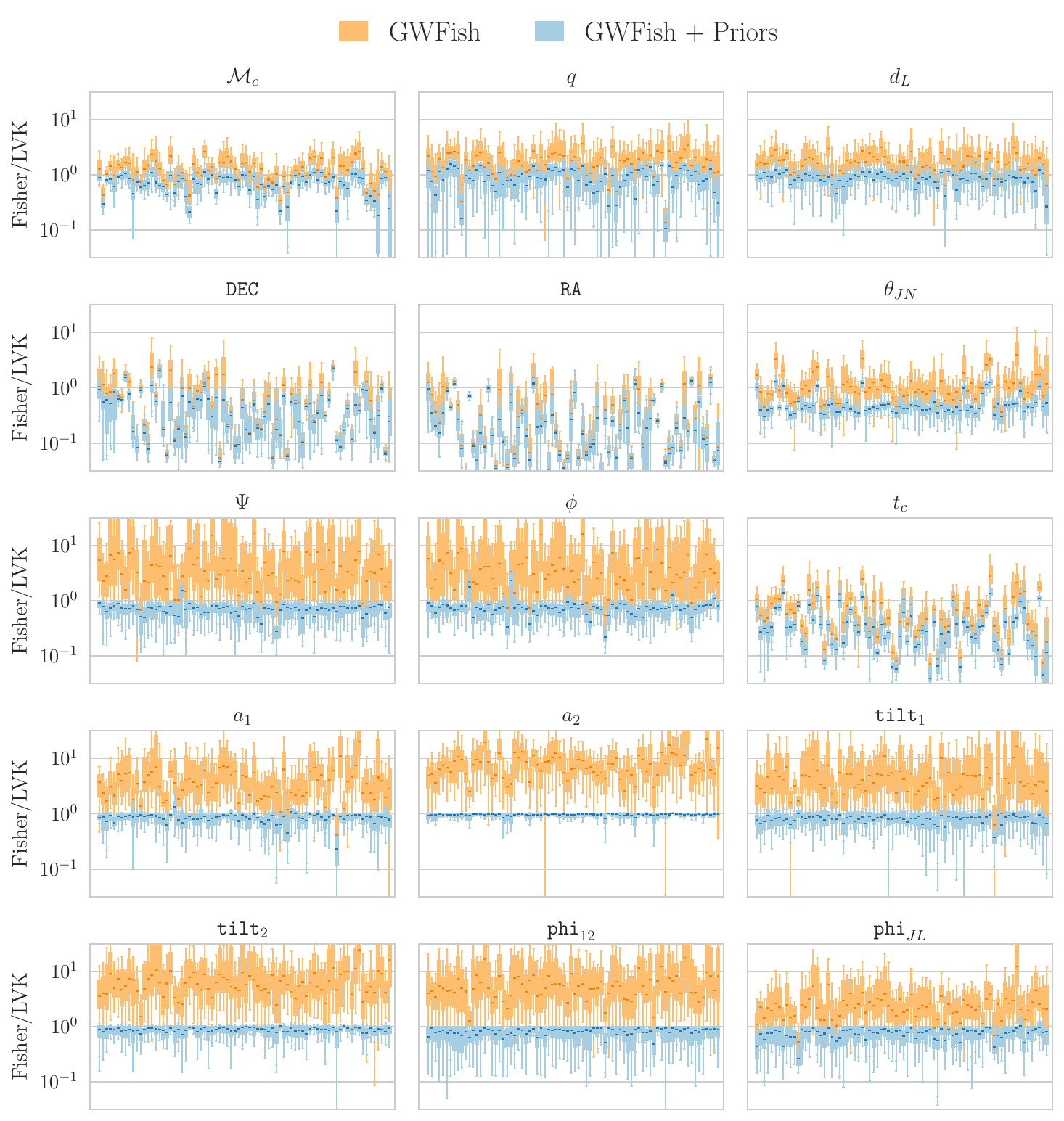}
    \caption{Distribution of ratios between the $90\%$ credible interval obtained from \texttt{GWFish} results and the full Bayesian parameter estimation (LVK) per each event and per each of the 15 parameters. Events are displayed along the $x-$axis, and each subplot refers to a different parameter. Results obtained without (with) priors are shown in \textit{orange} (\textit{blue}). The median value of each distribution per event is indicated as a horizontal line of a darker shade.}
    \label{fig:boxplot_summary}
\end{figure*}

\begin{figure*}[ht!]
    \centering
    \includegraphics[scale=0.7]{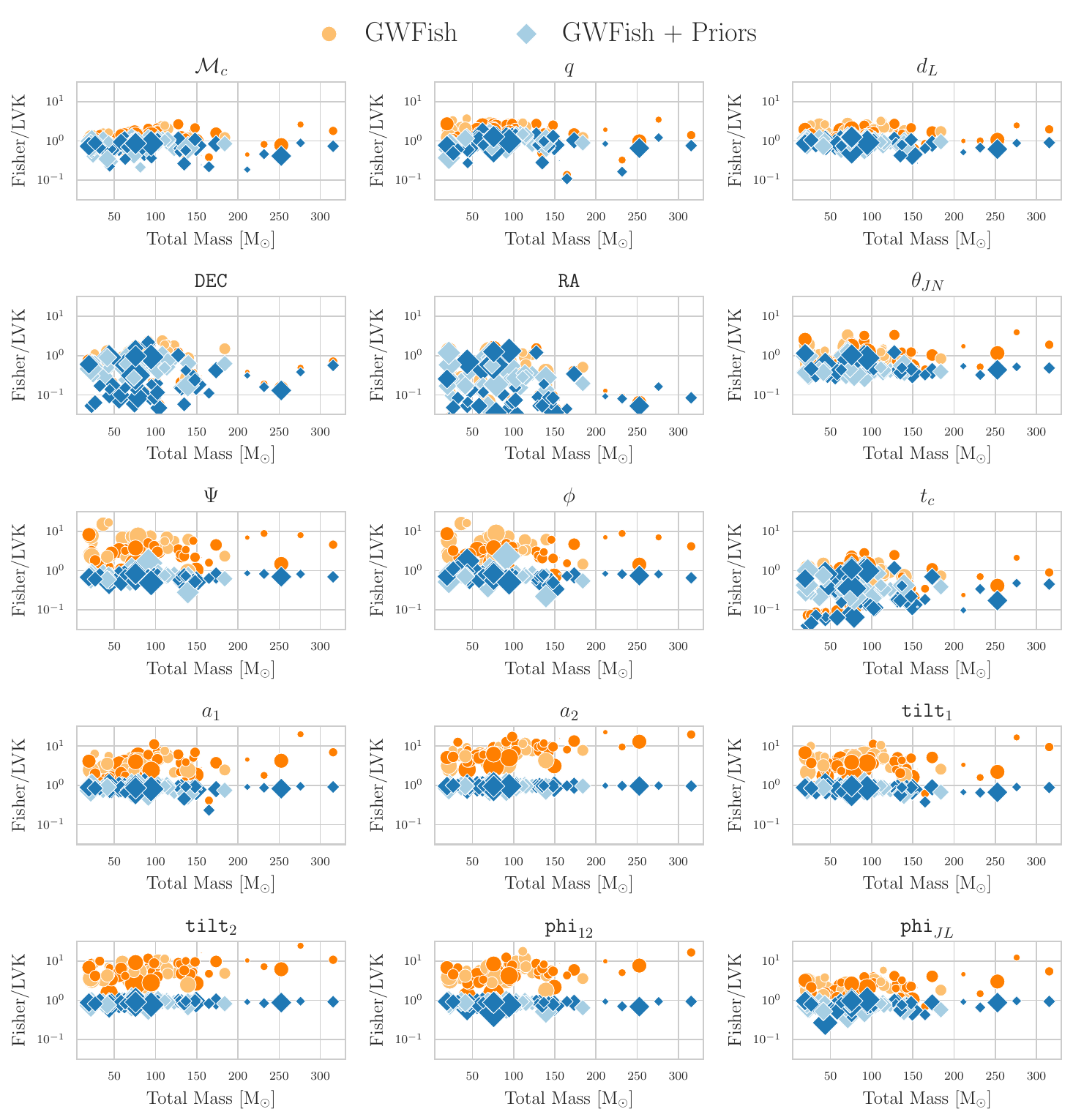}
    \caption{Median of the ratio between the $90\%$ credible interval obtained from \texttt{GWFish} results and full Bayesian parameter estimation (LVK). Results obtained without (with) priors are shown in \textit{orange} (\textit{blue}). The size of the markers is proportional to the median of the SNR of the network. Light (dark) shade refers to two (three) detectors observing the event.}
    \label{fig:median_summary}
\end{figure*}

\begin{figure*}[ht!]
    \centering
    \includegraphics[scale=0.7]{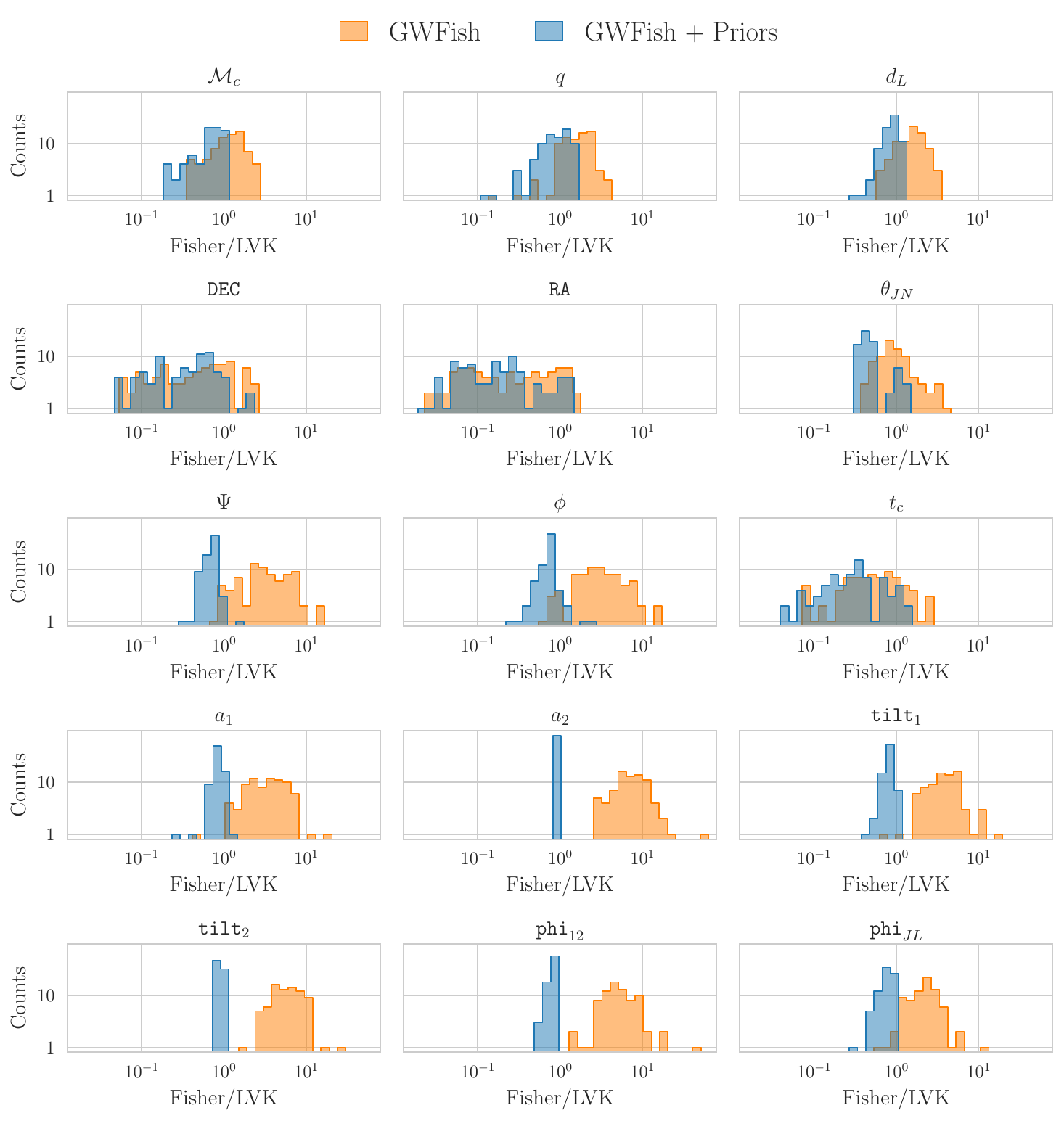}
    \caption{Histograms of the median of the ratio distribution between the 90\% credible interval obtained from \texttt{GWFish} results and full Bayesian parameter estimation (LVK). Results obtained without (with) priors are shown in \textit{orange} (\textit{blue}).}
    \label{fig:histograms_summary}
\end{figure*}

Our main results are summarized in Fig.~\ref{fig:boxplot_summary}, Fig.~\ref{fig:median_summary} and in Fig.~\ref{fig:histograms_summary}. In Fig.~\ref{fig:median_summary}, for each of the 15 parameters, we plot the ratio distribution between the $90\%$ credible interval we get from the Fisher analyses, without (in \textit{orange}) and with priors (in \textit{blue}), and the $90\%$ credible interval that was obtained with the full Bayesian approach. Events, along the $x-$axis are represented by boxplots. The median of the distribution for each event is highlighted as a horizontal line in a darker shade for both the analyses, with and without priors. In Fig.~\ref{fig:median_summary}, we plot the medians of the distributions as a function of the total mass of the binary (the median value in the detector frame). Furthermore, the size of the dots is proportional to the SNR of the signal (ranging in $[\sim 4, \sim 25]$), and the colour shade indicates whether the event was seen by 2 (lighter shade) or 3 detectors (darker shade). The SNR obtained in each of the detectors and for the network are listed in detail in Tab.~\ref{tab:events}. Fig.~\ref{fig:histograms_summary} shows the corresponding histograms of the medians of the $90\%$ credible interval ratio distributions per event for each parameter, with (in \textit{blue}) and without (in \textit{orange}) priors.
We are not evaluating the distance between our and the LVK results in a distributional sense using, for example, the Hellinger distance, as this leads to a case-by-case and, within one event, parameter-dependent results due to degeneracies and multimodality issues.
Here are the observations:
\begin{itemize}
    \item \textit{mass parameters}: The Fisher analysis tends to overestimate chirp mass and mass ratio errors. The match with LVK results is better for the chirp mass. When priors are included, the match becomes very good, and most of the bias of the error estimates (towards larger or smaller errors) is removed. 
    \item \textit{luminosity distance}: The Fisher analysis typically overestimates errors of the luminosity distance. The inclusion of priors leads to a good match with LVK results. When all three detectors are active (see Fig.~\ref{fig:median_summary_virgo} and discussion in Sec.~\ref{subsec:multi_modality} below), the constraint on the distance measurement is already done well by the Fisher analysis alone.
    \item \textit{angular parameters}: The inclusion of priors does not generally improve the error estimates of \texttt{RA}, \texttt{DEC}, and the difference between LVK results and Fisher analysis spans over two orders of magnitude. $\theta_{JN}$, $\Psi$ and $\phi$ show a slight improvement from priors. We address this specifically in a distinct Sec.~\ref{subsec:multi_modality} below.
    \item \textit{time of coalescence}: The error on $t_c$ is already well estimated with Fisher analysis alone. It is curious to observe that when priors are included in the \texttt{GWFish} analysis, the errors of the merger-time estimates can be significantly underestimated.
    \item \textit{spin parameters}: All the spin parameters are poorly constrained with Fisher analysis alone. Including priors allows us to reproduce the LVK results. Still, it should be noted that even the LVK results are likely prior driven since the signals reported in GWTC carried very little information about spins (with some exceptions). 
\end{itemize}

\subsubsection{Angular Parameters and Multi-Modality}
\label{subsec:multi_modality}
\begin{figure*}[ht!]
    \centering
    \includegraphics[scale=0.7]{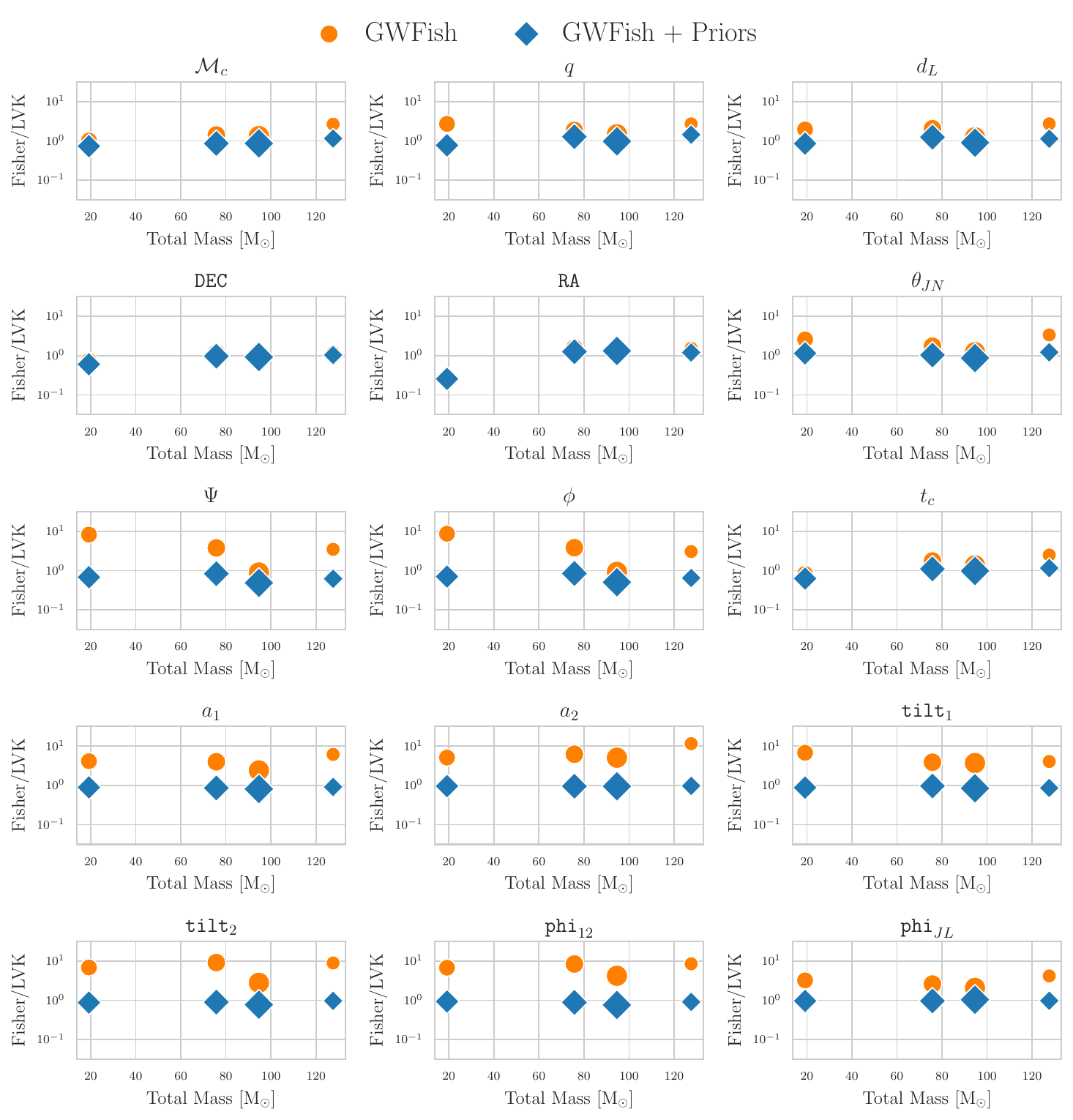}
    \caption{Same as in Fig.~\ref{fig:median_summary}, but here we show only those events detected with all three interferometers and SNR in all three detectors is $\gtrsim 4$. The network SNR for these events varies in the range between 9 and 19.}
    \label{fig:median_summary_virgo}
\end{figure*}

A full Bayesian analysis might encounter multi-modality issues. This happens especially when it comes to angular parameters (see, for example, the $\theta_{JN}$ parameter in Fig.~\ref{fig:posteriors_gw150914}). Moreover, it is more evident when fewer detectors are involved, and the SNR in each detector is low (below 4-5). Multi-modality cannot, by definition, be a feature coming out of Fisher's analysis, as it provides a Gaussian likelihood centred around the injected value. This explains the ample range (up to two orders of magnitude) of the ratio between the 90\% credible interval obtained with \texttt{GWFish} analyses and the LVK results from full Bayesian analysis as shown in Fig.~\ref{fig:boxplot_summary}, Fig.~\ref{fig:median_summary}, and Fig.~\ref{fig:histograms_summary}. Generally, the Fisher approximation tends to overestimate the uncertainties, as seen from parameters like the chirp mass, the mass ratio or the luminosity distance. However, when a parameter presents a multi-modal distribution in standard Bayesian analysis, the Fisher analysis provides poor error estimates as it represents a single mode. Therefore, the constraints from the Fisher analysis are likely to be tighter. To this scope, we selected a subsample of events, for which all 3 detectors were used and the SNR in each detector was around 4 or higher. This left us with 4 events, \texttt{GW190701}, \texttt{GW200202}, \texttt{GW200224} and \texttt{GW200311}. The results are shown in Fig.~\ref{fig:median_summary_virgo}. They confirm the general trend discussed before: Fisher analysis alone gives broader or comparable estimates when multi-modality is broken by having more detectors in the analysis. 

When it comes to sky localization with ET, multi-modality can also be suppressed by observing signals for long periods of time. For example, neutron-star binaries would be observed for hours up to a full day, and this would allow the use of the amplitude and phase modulations of the signal due to the rotation of the Earth to infer the source location.

\section{Discussion and Conclusions}
\label{sec:discussion}
In this work, we conducted a thorough analysis of the GWTC data to investigate the effectiveness of Fisher-matrix methods and the impact of priors. Our approach involved comparing the posterior samples generated by the LVK collaboration's official Bayesian analysis \citep{LIGOScientific:2018mvr, LIGOScientific:2020ibl, KAGRA:2021vkt} with the results obtained from our Fisher-matrix software, \texttt{GWFish} \cite{Dupletsa:2022scg}. We not only used Fisher analyses alone, but also incorporated a method to apply priors, thus ensuring that our posterior samples were comparable to those generated by the LVK collaboration on an equal basis. Since the true value of events' parameters is unknown, we analysed 30 different posterior samples realizations per event chosen among those provided by the LVK results.

Our research is motivated by the ongoing uncertainty surrounding the reliability of the Fisher-matrix approximation. In the field of next-generation GW observatories, a lot of effort is being devoted to understanding their capabilities and making population-level predictions. However, due to the computational complexity involved in this task, the Fisher-matrix approximation is currently the only feasible tool to carry out such analyses. Therefore, our study aims at investigating the Fisher-matrix method, identify its limitations, and propose potential improvements to enhance its accuracy.

Our findings on the validity of Fisher's approximation are centered around two main themes. The first concerns the interaction between SNR and correlations, multi-modality, and degeneracies, while the second focuses on the impact of prior information.

\begin{itemize}
    \item Fisher-matrix analyses are always introduced as an approximation valid in the high-SNR limit \citep{Vallisneri:2007ev, Rodriguez:2013mla}. The results of this work suggest, instead, that the accuracy of the Fisher analysis is mostly limited by model degeneracies and correlations among parameters, rather than the SNR. Specifically, degeneracies pose a major issue even if the SNR is high. It still holds that very low SNR Fisher matrix analyses provide less accurate estimates of parameter errors. We find that the Fisher analysis does well, with a trend to overestimate uncertainties, confirming the findings of \cite{Rodriguez:2013mla} (see Fig.~\ref{fig:boxplot_summary}, Fig.~\ref{fig:median_summary}, Fig.~\ref{fig:histograms_summary}). This is the case as long as the parameters are not degenerate and the Bayesian analysis does not output multi-modal posterior distributions. By construction, in fact, the Fisher analysis cannot represent multi-modal distributions. Therefore, when multi-modality comes into play, the Fisher analysis represents one of the modes (centered around the injected parameter value), and this explains why the uncertainties are sometimes better estimated with Fisher approximation than with Bayesian analysis (see the angular parameters and hence sky localization). Multi-modalities are not easy to deal with. Some insights on the issue are given in \citep{Roulet:2022kot} where ad-hoc reparametrizations are proposed.
    \item Adding priors improves the error estimates. Parameters like chirp mass, mass ratio, and luminosity distance benefit from including priors. Especially for luminosity distance, this assures that the posterior does not extend beyond the physical range to negative values. Priors' impact is even more evident for spin parameters that are poorly constrained by Fisher analyses alone. This also happens in the full Bayesian analysis, where some parameters, like spins, are as well prior-driven.
    On the other hand, regarding sky location and angular parameters in general, including prior, the error estimates might only slightly improve due to the inherent limitations of the Fisher analysis in handling multi-modal distributions.
    \item There is, in fact, an interplay among priors, correlations and SNR, especially when few detectors are involved. In this case, the results for angular parameters are prone to degeneracies. When the sky-localization suffers from multi-modality, the match between LVK results and \texttt{GWFish+Priors} is generally worse. Especially the sky-location errors produced by \texttt{GWFish+Priors} cannot be trusted in this case. As we have shown in Fig.~\ref{fig:median_summary_virgo}, however, degeneracies can be effectively broken by having a network of 3 detectors, with a non-negligible SNR in each of them (at least around 4). This not only improves the match between LVK and \texttt{GWFish+Priors} results with respect to sky localization but, in fact, improves the match with respect to all parameters. This highlights the role of priors in taming degeneracies, and it indicates how well \texttt{Fisher+Priors} methods work in the absence of important multi-modality in the posterior distribution. In contrast, error estimates of mass parameters and distance match the LVK results very well, even when the observation is made with only two detectors of the LVK network. 
\end{itemize}

We conclude that a Gaussian-likelihood approximation is a valid tool for providing estimates of parameter errors. However, it is crucial to note that this method has limitations. To corroborate the findings, more observations from 3-detectors need to be analyzed. 

This study supports the results obtained in all the science case studies carried out so far with Fisher-matrix methods \citep{Borhanian:2022czq, Ronchini:2022gwk, Iacovelli:2022bbs, Banerjee:2022gkv, Branchesi:2023mws, Loffredo:2024gmx}, with the exception of a few analyses that concern BBH localization with less than three detectors.

\acknowledgments
We thank the anonymous reviewer for their useful comments.

We thank Tito Dal Canton, Boris Goncharov, Biswajit Banerjee, and Riccardo Murgia for useful discussions and suggestions. We also thank Giancarlo Cella for providing the LVK internal review. 

U.~Dupletsa acknowledges partial financial support from MUR PRIN Grant No. 2022-Z9X4XS, funded by the European Union - Next Generation EU. 

F.~Santoliquido acknowledges financial support from the AHEAD2020 project (grant agreement n. 871158). 

K.~K.~Y.~Ng is supported by Miller Fellowship at Johns Hopkins University and Croucher Fellowship by the Croucher Foundation.

\appendix
\section{List of GWTC events}
\label{app:events}
In Table~\ref{tab:events} below, we have listed all the events that were observed during the first three observing runs, excluding the BNS event \texttt{GW170817} (see Tab.~\ref{tab:events_numbers} for a summary). We have highlighted the events that were used in our analysis while also specifying why some events were excluded. Our focus was on BBH sources analyzed with the waveform approximant \texttt{IMRPhenomXPHM}. Events that were not BBH or lacked data about analytical priors were excluded from the study.

\begin{longtable*}{c | c c c c c c c}
\caption{List of all observed events during the first 3 observing runs by the LVK Collaboration. The runs have been split into 4 parts, highlighted by different shades of blue, and indicating respectively the data-taking subdivision: O1 (3 events), O2 (7 events plus \texttt{GW170817}), O3-a (44 events) and O3-b (36 events). For each \texttt{eventID} we show if the event is a BBH and, in this case, if it has the analysis with the \texttt{IMRPhenomXPHM} waveform approximant. Furthermore, we specify if the official data report the analytic priors used for the analysis. Finally, we show what detectors have seen the event with the SNR of single detectors and of the network (calculated with \texttt{GWFish}). The reported SNR value is the median of the values obtained from the analysis over the 30 realizations. The last column signals if the event is included in our analysis (in case not the reason why is specified under the column \texttt{problem}).}\\
\label{tab:events}
\textbf{eventID} &\textbf{BBH} &\textbf{IMRPhenomXPHM} &\textbf{priors} &\textbf{network} &\textbf{\texttt{GWFish} SNR} &\textbf{problem} &\textbf{included} \\
& & & & &\textbf{detectors $\rightarrow$ network} & &
\\
\toprule
\rowcolor{Color1}
\texttt{GW150914\_095045}    &\cmark &\cmark &\cmark &\scriptsize{[H1, L1]} &\scriptsize{$[19.4, 13.8 \rightarrow 23.8]$} &- &\cmark 
\\
\texttt{GW151012\_095443}    &\cmark &\cmark &\cmark &\scriptsize{[H1, L1]} &\scriptsize{$[6.4, 6.0 \rightarrow 8.7]$} &- &\cmark 
\\
\rowcolor{Color1}
\texttt{GW151226\_033853}    &\cmark &\cmark &\cmark &\scriptsize{[H1, L1]} &\scriptsize{$[9.5, 7.2 \rightarrow 11.8]$} &- &\cmark 
\\
\hline
\hline
\rowcolor{Color2}
\texttt{GW170104\_101158}    &\cmark &\cmark &\cmark &\scriptsize{[H1, L1]} &\scriptsize{$[8.8, 9.6 \rightarrow 12.9]$} &- &\cmark     
\\
\texttt{GW170608\_020116}    &\cmark &\cmark &\xmark &\scriptsize{[H1, L1]} &- &\scriptsize{no priors} &\xmark     
\\
\rowcolor{Color2}
\texttt{GW170729\_185629} &\cmark &\cmark &\cmark &\scriptsize{[H1, L1]} &\scriptsize{$[5.4, 7.6 \rightarrow 9.3]$} &- &\cmark     
\\
\texttt{GW170809\_082821}    &\cmark &\cmark &\cmark &\scriptsize{[H1, L1]} &\scriptsize{$[1.9, 10.2 \rightarrow 10.5]$} &- &\cmark     
\\
\rowcolor{Color2}
\texttt{GW170814\_103043}    &\cmark &\cmark &\cmark &\scriptsize{[H1, L1, V1]} &\scriptsize{$[3.4, 13.3, 4.2 \rightarrow 14.4]$} &- &\cmark   
\\
\texttt{GW170818\_022509}   &\cmark &\cmark &\cmark &\scriptsize{[H1, L1, V1]} &\scriptsize{$[3.2, 9.2, 4.2 \rightarrow 10.7]$} &- &\cmark 
\\
\rowcolor{Color2}
\texttt{GW170823\_131358}    &\cmark &\cmark &\cmark &\scriptsize{[H1, L1]} &\scriptsize{$[6.6, 8.9 \rightarrow 11.3]$} &- &\cmark
\\
\hline
\hline
\rowcolor{Color3}
\texttt{GW190403\_051519}    &\cmark &\cmark &\cmark &\scriptsize{[H1, L1, V1]} &\scriptsize{$[3.8, 5.1, 1.1 \rightarrow 6.4]$} &- &\cmark
\\
\texttt{GW190408\_181802}    &\cmark &\cmark &\cmark &\scriptsize{[H1, L1, V1]} &\scriptsize{$[9.5, 10.7, 2.6 \rightarrow 14.6]$} &- &\cmark
\\
\rowcolor{Color3}
\texttt{GW190412\_053044}    &\cmark &\cmark &\cmark &\scriptsize{[H1, L1, V1]} &\scriptsize{$[9.9, 15.4, 3.3 \rightarrow 18.8]$} &- &\cmark
\\
\texttt{GW190413\_052954}    &\cmark &\cmark &\cmark &\scriptsize{[H1, L1, V1]} &\scriptsize{$[5.5, 8.8, 1.4 \rightarrow 10.4]$} &- &\cmark
\\
\rowcolor{Color3}
\texttt{GW190413\_134308}    &\cmark &\cmark &\cmark &\scriptsize{[H1, L1, V1]} &\scriptsize{$[4.0, 6.1, 2.1 \rightarrow 7.7]$} &- &\cmark
\\
\texttt{GW190421\_213856}    &\cmark &\cmark &\cmark &\scriptsize{[H1, L1]} &\scriptsize{$[7.5, 6.7 \rightarrow 10.1]$} &- &\cmark
\\
\rowcolor{Color3}
\texttt{GW190425\_081805}   &\xmark &- &- &- &- &\scriptsize{no BBH} &\xmark 
\\
\texttt{GW190426\_190642}    &\cmark &\cmark &\cmark &\scriptsize{[H1, L1, V1]} &\scriptsize{$[2.7, 6.6, 1.4 \rightarrow 7.3]$} &- &\cmark
\\
\rowcolor{Color3}
\texttt{GW190503\_185404}    &\cmark &\cmark &\cmark &\scriptsize{[H1, L1, V1]} &\scriptsize{$[5.7, 7.2, 2.3 \rightarrow 9.6]$} &- &\cmark
\\
\texttt{GW190512\_180714}    &\cmark &\cmark &\cmark &\scriptsize{[H1, L1, V1]} &\scriptsize{$[6.2, 10.1, 1.4 \rightarrow 12.0]$} &- &\cmark
\\
\rowcolor{Color3}
\texttt{GW190513\_205428}    &\cmark &\cmark &\cmark &\scriptsize{[H1, L1, V1]} &\scriptsize{$[7.4, 8.2, 1.8 \rightarrow 11.3]$} &- &\cmark
\\
\texttt{GW190514\_065416}    &\cmark &\cmark &\cmark &\scriptsize{[H1, L1]} &\scriptsize{$[4.5, 5.1 \rightarrow 6.6]$} &- &\cmark
\\
\rowcolor{Color3}
\texttt{GW190517\_055101}    &\cmark &\cmark &\cmark &\scriptsize{[H1, L1, V1]} &\scriptsize{$[4.3, 7.1, 3.4 \rightarrow 9.2]$} &- &\cmark
\\
\texttt{GW190519\_153544}    &\cmark &\cmark &\cmark &\scriptsize{[H1, L1, V1]} &\scriptsize{$[8.4, 12.5, 1.6 \rightarrow 14.7]$} &- &\cmark
\\
\rowcolor{Color3}
\texttt{GW190521\_030229}    &\cmark &\cmark &\cmark &\scriptsize{[H1, L1, V1]} &\scriptsize{$[6.2, 11.6, 1.9 \rightarrow 13.7]$} &- &\cmark
\\
\texttt{GW190521\_074359}    &\cmark &\cmark &\cmark &\scriptsize{[H1, L1]} &\scriptsize{$[11.5, 20.7\rightarrow 23.4]$} &- &\cmark
\\
\rowcolor{Color3}
\texttt{GW190527\_092055}    &\cmark &\cmark &\cmark &\scriptsize{[H1, L1]} &\scriptsize{$[3.5, 5.2 \rightarrow 6.8]$} &- &\cmark
\\
\texttt{GW190602\_175927}    &\cmark &\cmark &\cmark &\scriptsize{[H1, L1, V1]} &\scriptsize{$[3.0, 9.7, 2.3 \rightarrow 10.6]$} &- &\cmark
\\
\rowcolor{Color3}
\texttt{GW190620\_030421}    &\cmark &\cmark &\cmark &\scriptsize{[L1, V1]} &\scriptsize{$[10.5, 2.3 \rightarrow 10.7]$} &- &\cmark
\\
\texttt{GW190630\_185205}    &\cmark &\cmark &\cmark &\scriptsize{[L1, V1]} &\scriptsize{$[14.4, 5.1 \rightarrow 15.3]$} &- &\cmark
\\
\rowcolor{Color3}
\texttt{GW190701\_203306}    &\cmark &\cmark &\cmark &\scriptsize{[H1, L1, V1]} &\scriptsize{$[4.3, 6.3, 4.3 \rightarrow 9.0]$} &- &\cmark
\\
\texttt{GW190706\_222641}    &\cmark &\cmark &\cmark &\scriptsize{[H1, L1]} &\scriptsize{$[6.3, 7.3 \rightarrow 9.8]$} &- &\cmark
\\
\rowcolor{Color3}
\texttt{GW190707\_093326}    &\cmark &\cmark &\xmark &\scriptsize{[H1, L1]} &- &\scriptsize{no priors} &\xmark
\\
\texttt{GW190708\_232457}    &\cmark &\cmark &\cmark &\scriptsize{[L1, V1]} &\scriptsize{$[12.5, 2.5 \rightarrow 12.8]$} &- &\cmark
\\
\rowcolor{Color3}
\texttt{GW190719\_215514}    &\cmark &\cmark &\cmark &\scriptsize{[H1, L1]} &\scriptsize{$[4.4, 5.1 \rightarrow 6.8]$} &- &\cmark
\\
\texttt{GW190720\_000836}    &\cmark &\cmark &\xmark &\scriptsize{[H1, L1, V1]} &- &\scriptsize{no priors} &\xmark
\\
\rowcolor{Color3}
\texttt{GW190725\_174728}    &\cmark &\cmark &\xmark &\scriptsize{[H1, L1, V1]} &- &\scriptsize{no priors} &\xmark
\\
\texttt{GW190727\_060333}    &\cmark &\cmark &\cmark &\scriptsize{[H1, L1, V1]} &\scriptsize{$[6.8, 8.5, 2.9 \rightarrow 11.3]$} &- &\cmark
\\
\rowcolor{Color3}
\texttt{GW190728\_064510}    &\cmark &\cmark &\xmark &\scriptsize{[H1, L1, V1]} &- &\scriptsize{no priors} &\xmark
\\
\texttt{GW190731\_140936}    &\cmark &\cmark &\cmark &\scriptsize{[H1, L1]} &\scriptsize{$[4.1, 5.0 \rightarrow 6.9]$} &- &\cmark
\\
\rowcolor{Color3}
\texttt{GW190803\_022701}    &\cmark &\cmark &\cmark &\scriptsize{[H1, L1, V1]} &\scriptsize{$[5.0, 6.0, 1.2 \rightarrow 7.8]$} &- &\cmark
\\
\texttt{GW190805\_211137}    &\cmark &\cmark &\cmark &\scriptsize{[H1, L1, V1]} &\scriptsize{$[3.8, 6.1, <1 \rightarrow 7.1]$} &- &\cmark
\\
\rowcolor{Color3}
\texttt{GW190814\_211039}    &\cmark &\cmark &\xmark &\scriptsize{[H1, L1, V1]} &- &\scriptsize{no priors} &\xmark
\\
\texttt{GW190828\_063405}    &\cmark &\cmark &\cmark &\scriptsize{[H1, L1, V1]} &\scriptsize{$[9.5, 11.7, 1.8 \rightarrow 15.2]$} &- &\cmark
\\
\rowcolor{Color3}
\texttt{GW190828\_065509}    &\cmark &\cmark &\cmark &\scriptsize{[H1, L1, V1]} &\scriptsize{$[6.1, 7.0, 1.4 \rightarrow 9.2]$} &- &\cmark
\\
\texttt{GW190910\_112807}    &\cmark &\cmark &\cmark &\scriptsize{[L1, V1]} &\scriptsize{$[13.3, 2.5 \rightarrow 13.8]$} &- &\cmark
\\
\rowcolor{Color3}
\texttt{GW190915\_235702}    &\cmark &\cmark &\cmark &\scriptsize{[H1, L1, V1]} &\scriptsize{$[9.4, 7.5, 2.3 \rightarrow 12.3]$} &- &\cmark
\\
\texttt{GW190916\_200658}    &\cmark &\cmark &\cmark &\scriptsize{[H1, L1, V1]} &\scriptsize{$[4.5, 5.2, 1.1 \rightarrow 6.9]$} &- &\cmark
\\
\rowcolor{Color3}
\texttt{GW190917\_114630}    &\cmark &\cmark &\xmark &\scriptsize{[H1, L1, V1]} &- &\scriptsize{no priors} &\xmark
\\
\texttt{GW190924\_021846}    &\cmark &\cmark &\xmark &\scriptsize{[H1, L1, V1]} &- &\scriptsize{no priors} &\xmark
\\
\rowcolor{Color3}
\texttt{GW190925\_232845}    &\cmark &\cmark &\cmark &\scriptsize{[H1, V1]} &\scriptsize{$[6.3, 4.4 \rightarrow 7.4]$} &- &\cmark
\\
\texttt{GW190926\_050336}    &\cmark &\cmark &\cmark &\scriptsize{[H1, L1, V1]} &\scriptsize{$[4.0, 5.0, 1.0 \rightarrow 6.9]$} &- &\cmark
\\
\rowcolor{Color3}
\texttt{GW190929\_012149}    &\cmark &\cmark &\cmark &\scriptsize{[H1, L1, V1]} &\scriptsize{$[5.7, 6.5, <1 \rightarrow 8.9]$} &- &\cmark
\\
\texttt{GW190930\_133541}    &\cmark &\cmark &\cmark &\scriptsize{[H1, L1]} &\scriptsize{$[3.9, 6.5 \rightarrow 8.2]$} &- &\cmark
\\
\hline
\hline
\rowcolor{Color4}
\texttt{GW191103\_012549}    &\cmark &\cmark &\cmark &\scriptsize{[H1, L1]} &\scriptsize{$[5.3, 6.0 \rightarrow 8.0]$} &- &\cmark
\\
\texttt{GW191105\_143521}    &\cmark &\cmark &\cmark &\scriptsize{[H1, L1, V1]} &\scriptsize{$[5.2, 7.0, 1.3 \rightarrow 9.0]$} &- &\cmark
\\
\rowcolor{Color4}
\texttt{GW191109\_010717}    &\cmark &\cmark &\cmark &\scriptsize{[H1, L1]} &\scriptsize{$[8.7, 13.1 \rightarrow 15.6]$} &- &\cmark
\\
\texttt{GW191113\_071753}    &\cmark &\cmark &\cmark &\scriptsize{[H1, L1, V1]} &\scriptsize{$[4.1, 5.1, 1.1 \rightarrow 6.7]$} &- &\cmark
\\
\rowcolor{Color4}
\texttt{GW191126\_115259}    &\cmark &\cmark &\cmark &\scriptsize{[H1, L1]} &\scriptsize{$[4.8, 5.8 \rightarrow 7.5]$} &- &\cmark
\\
\texttt{GW191127\_050227}    &\cmark &\cmark &\cmark &\scriptsize{[H1, L1, V1]} &\scriptsize{$[5.9, 6.1, 2.5 \rightarrow 8.8]$} &- &\cmark
\\
\rowcolor{Color4}
\texttt{GW191129\_134029}    &\cmark &\cmark &\cmark &\scriptsize{[H1, L1]} &\scriptsize{$[7.4, 9.7 \rightarrow 11.9]$} &- &\cmark
\\
\texttt{GW191204\_110529}    &\cmark &\cmark &\cmark &\scriptsize{[H1, L1, V1]} &\scriptsize{$[5.3, 6.3 \rightarrow 8.4]$} &- &\cmark
\\
\rowcolor{Color4}
\texttt{GW191204\_171526}    &\cmark &\cmark &\cmark &\scriptsize{[H1, L1]} &\scriptsize{$[9.7, 12.9 \rightarrow 16.2]$} &- &\cmark
\\
\texttt{GW191215\_223052}    &\cmark &\cmark &\cmark &\scriptsize{[H1, L1, V1]} &\scriptsize{$[6.5, 7.9, 1.5 \rightarrow 10.8]$} &- &\cmark
\\
\rowcolor{Color4}
\texttt{GW191216\_213338}    &\cmark &\cmark &\cmark &\scriptsize{[H1, V1]} &\scriptsize{$[16.3, 2.5 \rightarrow 16.4]$} &- &\cmark
\\
\texttt{GW191219\_163120}    &\xmark &- &- &- &- &\scriptsize{no BBH} &\xmark
\\
\rowcolor{Color4}
\texttt{GW191222\_033537}    &\cmark &\cmark &\cmark &\scriptsize{[H1, L1]} &\scriptsize{$[7.4, 8.7 \rightarrow 11.4]$} &- &\cmark
\\
\texttt{GW191230\_180458}    &\cmark &\cmark &\cmark &\scriptsize{[H1, L1, V1]} &\scriptsize{$[5.5, 6.2, 1.4 \rightarrow 8.5]$} &- &\cmark
\\
\rowcolor{Color4}
\texttt{GW200105\_162426}    &\xmark &- &- &- &- &\scriptsize{no BBH} &\xmark
\\
\texttt{GW200112\_155838}    &\cmark &\cmark &\cmark &\scriptsize{[L1, V1]} &\scriptsize{$[18.2, 3.4 \rightarrow 18.6]$} &- &\cmark
\\
\rowcolor{Color4}
\texttt{GW200115\_042309}    &\xmark &- &- &- &- &\scriptsize{no BBH} &\xmark
\\
\texttt{GW200128\_022011}     &\cmark &\cmark &\cmark &\scriptsize{[H1, L1]} &\scriptsize{$[6.7, 7.3 \rightarrow 9.9]$} &- &\cmark
\\
\rowcolor{Color4}
\texttt{GW200129\_065458}    &\cmark &\cmark &\cmark &\scriptsize{[H1, L1, V1]} &\scriptsize{$[13.2, 20.3, 2.9 \rightarrow 24.7]$} &- &\cmark
\\
\texttt{GW200202\_154313}    &\cmark &\cmark &\cmark &\scriptsize{[H1, L1, V1]} &\scriptsize{$[7.7, 8.9, 4.1 \rightarrow 12.6]$} &- &\cmark
\\
\rowcolor{Color4}
\texttt{GW200208\_130117}    &\cmark &\cmark &\cmark &\scriptsize{[H1, L1, V1]} &\scriptsize{$[2.6, 6.9, 4.5 \rightarrow 8.6]$} &- &\cmark
\\
\texttt{GW200208\_222617}    &\cmark &\cmark &\cmark &\scriptsize{[H1, L1, V1]} &\scriptsize{$[4.5, 5.1, 1.2 \rightarrow 7.0]$} &- &\cmark
\\
\rowcolor{Color4}
\texttt{GW200209\_085452}    &\cmark &\cmark &\cmark &\scriptsize{[H1, L1, V1]} &\scriptsize{$[6.3, 5.8, 1.6 \rightarrow 8.9]$} &-  &\cmark
\\
\texttt{GW200210\_092254}    &\cmark &\cmark &\cmark &\scriptsize{[H1, L1, V1]} &\scriptsize{$[4.8, 5.9, 1.0 \rightarrow 7.9]$} &- &\cmark
\\
\rowcolor{Color4}
\texttt{GW200216\_220804}    &\cmark &\cmark &\cmark &\scriptsize{[H1, L1, V1]} &\scriptsize{$[5.1, 4.3, 1.6 \rightarrow 6.9]$} &- &\cmark
\\
\texttt{GW200219\_094415}    &\cmark &\cmark &\cmark &\scriptsize{[H1, L1, V1]} &\scriptsize{$[6.2, 7.5, 1.1 \rightarrow 9.7]$}  &- &\cmark
\\
\rowcolor{Color4}
\texttt{GW200220\_061928}    &\cmark &\cmark &\cmark &\scriptsize{[H1, L1, V1]} &\scriptsize{$[2.5, 5.2, 1.6 \rightarrow 5.8]$} &- &\cmark
\\
\texttt{GW200220\_124850}    &\cmark &\cmark &\cmark &\scriptsize{[H1, L1]} &\scriptsize{$[5.1, 5.4 \rightarrow 7.7]$} &- &\cmark
\\
\rowcolor{Color4}
\texttt{GW200224\_222234}    &\cmark &\cmark &\cmark &\scriptsize{[H1, L1, V1]} &\scriptsize{$[12.8, 13.3, 4.4 \rightarrow 19.1]$} &- &\cmark
\\
\texttt{GW200225\_060421}    &\cmark &\cmark &\cmark &\scriptsize{[H1, L1]} &\scriptsize{$[8.8, 7.6 \rightarrow 11.7]$} &- &\cmark
\\
\rowcolor{Color4}
\texttt{GW200302\_015811}    &\cmark &\cmark &\cmark &\scriptsize{[H1, V1]} &\scriptsize{$[9.9, 2.5 \rightarrow 10.1]$} &- &\cmark
\\
\texttt{GW200306\_093714}    &\cmark &\cmark &\cmark &\scriptsize{[H1, L1, V1]} &\scriptsize{$[4.8, 4.9, 1.4 \rightarrow 7.2]$} &- &\cmark
\\
\rowcolor{Color4}
\texttt{GW200308\_173609}    &\cmark &\cmark &\cmark &\scriptsize{[H1, L1, V1]} &\scriptsize{$[2.8, 3.5, 1.2 \rightarrow 4.8]$} &- &\cmark
\\
\texttt{GW200311\_115853}    &\cmark &\cmark &\cmark &\scriptsize{[H1, L1, V1]} &\scriptsize{$[10.3, 9.3, 5.6 \rightarrow 15.0]$} &- &\cmark
\\
\rowcolor{Color4}
\texttt{GW200316\_215756}    &\cmark &\cmark &\cmark &\scriptsize{[H1, L1, V1]} &\scriptsize{$[5.4, 7.3, 2.5 \rightarrow 9.4]$} &- &\cmark
\\
\texttt{GW200322\_091133}    &\cmark &\cmark &\cmark &\scriptsize{[H1, L1, V1]} &\scriptsize{$[2.5, 2.1, <1 \rightarrow 3.6]$} &-  &\cmark
\\
\bottomrule

\end{longtable*}

\section{Prior probability selection in GW Analysis}
\label{app:choosing_priors}
In the following, we present an overview of the standard choices made in GW analysis for selecting priors. The LVK papers \cite{LIGOScientific:2018mvr, LIGOScientific:2020ibl, KAGRA:2021vkt} adhere to these standard choices. Given the absence of a common reference, we are providing a comprehensive summary of the core mathematical principles that underlie the selection of priors.

\subsection{\texttt{Uniform Prior}: the uninformative choice}
Most of the parameters characterizing a GW event (see Tab \ref{tab:priors}) have an uninformative uniform prior, meaning that given a physical range of the parameter, any value in the interval is equally probable. The probability density for a uniform distribution is given by:
\begin{equation}
    \label{eq:uniform_prior}
    \pi(\theta) = \frac{1}{\theta_{\rm max} - \theta_{\rm min}}
\end{equation}
where $ [\theta_{\rm min}, \theta_{\rm max}] $ is the range of the parameter $\theta$. The probability density is zero outside this interval, meaning values out of range are excluded from consideration. Choosing this prior implies that no prior knowledge favours one value over another within the interval.

\subsection{${\mathcal{M}}_c$ and $q$: \texttt{Uniform in component masses prior}}
LVK analyses assume a uniform in component masses prior, meaning that the joint probability on $m_1$ and $m_2$, $\pi(m_1, m_2)$, is constant. Since we are getting likelihood samples in ${\mathcal{M}}_{\rm chirp}$ and $q$, it is necessary to reweight these samples by the determinant of the Jacobian transformation to component masses. This ensures that the probability remains uniform in the $m_1$-$m_2$ plane:
\begin{equation}
    \label{eq:masses_jac}
    \pi(\mathcal{M}_{\rm chirp},q) = \pi( m_1, m_2) |J|,
\end{equation}
where the following transformation matrix gives $J$, the Jacobian:
\begin{equation}
    J =
\begin{pmatrix}
\frac{\partial m_1}{\partial \mathcal{M}_{\rm chirp}} &\frac{\partial m_1}{\partial q}\\
\frac{\partial m_2}{\partial \mathcal{M}_{\rm chirp}} &\frac{\partial m_2}{\partial q}
\end{pmatrix}.
\end{equation}
The determinant of the Jacobian, $J$, is:
\begin{equation}
    |J| = \mathcal{M}_{\rm chirp}q^{-6/5}(1+q)^{2/5}
\end{equation}
given that:
\begin{equation}
    \begin{cases}
m_1 &= \mathcal{M}_{\rm chirp}q^{-3/5}(1+q)^{1/5}\\
m_2 &= \mathcal{M}_{\rm chirp}q^{2/5}(1+q)^{1/5}
\end{cases}
\end{equation}

\subsection{Angular parameters: \texttt{Uniform on a Sphere} prior}
\label{subsec:angles}
While a \texttt{Uniform} prior might be the unbiased choice in many contexts, we must consider the spherical geometry when dealing with angular parameters describing positions on a sphere.
To ensure a uniform distribution on a sphere, thus isotropy, of certain angular parameters, the right priors are the ones that are uniform in either \texttt{Sine} or \texttt{Cosine}, depending on the physical range of the angular parameter in consideration.

\subsubsection{\texttt{RA} and \texttt{DEC}: a reference example}
We choose two well-known parameters to explain how angular priors work: the position in the sky given by \texttt{RA} and \texttt{DEC} (see also Fig.~\ref{fig:ra_dec}). The azimuthal angle \texttt{RA} ranges in $[0, 2\pi]$ with the zero reference given by the meridian passing through the vernal equinox. The polar angle \texttt{DEC} ranges in $[-\pi/2, +\pi/2]$, with zero value at the equator, positive values above and negative below. A uniform distribution on a sphere implies an equal probability of finding an object for any region on the sphere of equal area. This differs from individually considering a uniform distribution for \texttt{RA} and \texttt{DEC}. The reason is that lines of constant \texttt{DEC}, for example, are circles of different sizes, the ones near the pole smaller than the ones near the equator. A uniform prior in \texttt{DEC} would lead to a higher concentration of points near the poles.

The correct approach is to consider the area element on the sphere:
\begin{equation}
    \label{eq:spherical_area_element}
    \begin{aligned}
    \mathrm{d}A &= \sin\left(\frac{\pi}{2} - \texttt{DEC}\right)\mathrm{d}\texttt{DEC}\mathrm{d}\texttt{RA}\\
    &= \cos(\texttt{DEC})\mathrm{d}\texttt{DEC}\mathrm{d}\texttt{RA}\\
    &=\mathrm{d}\sin(\texttt{DEC})\mathrm{d}\texttt{RA}
    \end{aligned}
\end{equation}
where we have accounted for the fact that \texttt{DEC} is not measured starting from the north pole. 
From Eq.~\ref{eq:spherical_area_element}, we can see that if  \texttt{RA} is uniformly distributed in $[0, 2\pi]$, for \texttt{DEC} the distribution is uniform in \texttt{Sine(DEC)}:
\begin{equation}
    \pi(\sin(\texttt{DEC})) \propto \text{const}
\end{equation}
where the constant can be found by integrating in the \texttt{DEC} range and obtaining $\text{const}=1/2$. We can change variable and directly obtain the prior probability density for declination angle parameter:
\begin{equation}
\label{eq:dec_prior}
\begin{aligned}
\pi(\texttt{DEC}) &= \pi(\sin(\texttt{DEC}))\left|\frac{\mathrm{d}\sin(\texttt{DEC})}{\mathrm{d}\texttt{DEC}}\right|\\
&= \frac{1}{2}\cos(\texttt{DEC}) \quad \text{for} \quad -\frac{\pi}{2}\le \text{\texttt{DEC}} \le +\frac{\pi}{2}
\end{aligned}
\end{equation}

\subsubsection{Angular priors in synthesis}
In general, azimuthal angles, like \texttt{RA}, follow a uniform distribution in $[0, 2\pi]$. This holds for the phase $\phi$, the spin parameters $\texttt{phi}_{12}$ and $\texttt{phi}_{JL}$, and the polarization angle $\Psi$. The prior on the polarization angle $\Psi$, though, is uniform in $[0, \pi]$. The range is not $[0, 2\pi]$ due to the symmetry of the two GW polarizations, $h_{+}$ and $h_{\times}$. In fact, rotating the polarizations by $\pi$ does not affect the physical configuration of the system, as the polarization angle enters the signal as $\sin(2\Psi)$ or $\cos(2\Psi)$\footnote{The dependence on the polarization angle enters the antenna pattern functions:
\[
\begin{aligned}
F_{+}(\texttt{RA}, \texttt{DEC}, \Psi) &= F_{+,0}(\texttt{RA}, \texttt{DEC})\cos(2\Psi) - F_{\times, 0}(\texttt{RA}, \texttt{DEC})\sin(2\Psi)\\
F_{\times}(\texttt{RA}, \texttt{DEC}, \Psi) &= F_{+,0}(\texttt{RA}, \texttt{DEC})\sin(2\Psi) + F_{\times, 0}(\texttt{RA}, \texttt{DEC})\cos(2\Psi)\\
\end{aligned}
\]


where the $F_{+,0}(\texttt{RA}, \texttt{DEC}), F_{\times,0}(\texttt{RA}, \texttt{DEC})$ are the detector's response to the plus and cross polarizations depending only on the sky position of the source} without considering $\Psi$. 
Taking the range of the polarization angle to be between $[0, \pi]$ allows to avoid redundancy in parameter space, which is particularly useful in Bayesian analysis. 

The parameters $\theta_{JN}$, $\texttt{tilt}_1$ and $\texttt{tilt}_2$, instead, follow a zenith-like angular distribution as \texttt{DEC}. All three are uniformly distributed on the sphere in the range $[0, \pi]$. This means their Cosine is uniformly distributed. For example, the prior distribution for the $\theta_{JN}$ parameter is given by:
\begin{equation}
    \label{eq:sine_prior}
    \pi(\theta_{JN}) = \frac{1}{2} \sin(\theta_{JN}) \quad \text{for} \quad 0\le \theta_{JN} \le \pi
\end{equation}
which can be obtained in a analogous way as Eq.~\ref{eq:dec_prior} starting from $p(\cos(\texttt{DEC})) \propto \text{const}$.

\begin{figure}
\includegraphics[]{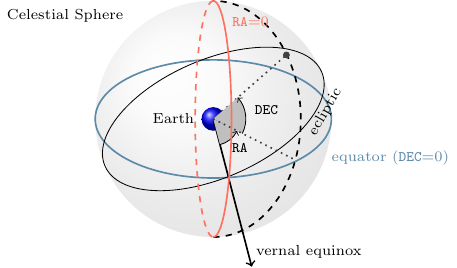}
\caption{Schematic representation of angular parameters that describe the position of the source in the sky. \texttt{RA} and \texttt{DEC} correspond, respectively, to the longitude and latitude on Earth, but in reference to the celestial sphere. The zero reference for \texttt{DEC} is the Earth's equator projected onto the celestial sphere. Instead, the zero reference for \texttt{RA} is the meridian on the celestial sphere passing through the intersection of the equator line and the ecliptic (vernal equinox). See also Tab.~\ref{tab:params_description}.}
\label{fig:ra_dec}
\end{figure}

\begin {figure}
\includegraphics[]{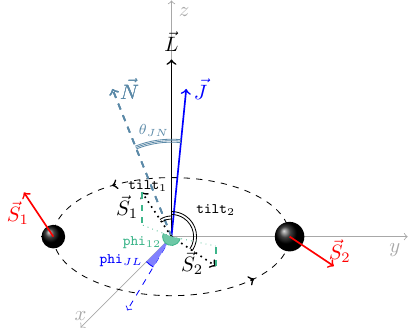}
\caption{Schematic view of relevant angular parameters. $\vec{N}$ is the line of sight, connecting the center of Earth to the center of the elliptic orbit. $\vec{L}$ is the orbital angular momentum of the binary, perpendicular to the orbital plane. $\vec{S}_1$ and $\vec{S}_2$ are spin components, with dimensionless magnitude proportional to $a_1\propto|\vec{S}_1|$ and $a_2 \propto |\vec{S}_2|$. \texttt{tilt\_1} and \texttt{tilt\_2} are the angles between $\vec{L}$ and single spin vectors, respectively. \texttt{phi\_JL} is the azimuthal angle of $\vec{L}$ around the total momentum $\vec{J}$ (sum of $\vec{L}$, individual spins $\vec{S}_1$, $\vec{S}_2$ and General Relativity corrections). It defines where the orbital angular momentum $\vec{L}$ lies on its precession cone around $\vec{J}$. \texttt{phi\_12} is the angle between the projections of spin vectors on the orbital plane. Note that \texttt{phi\_JL} and \texttt{phi\_12} live in two different planes: the first in the plane perpendicular to $\vec{J}$ and the second in the orbital plane, perpendicular to $\vec{L}$. In the case of aligned spins, the two planes coincide. Last, \texttt{theta\_JN} is the angle between the line of sight $\vec{N}$ and the total angular momentum $\vec{J}$. $\texttt{theta\_JN}$ reduces to the inclination angle $\iota$, when the spins of the binary components are aligned to the orbital angular momentum. See also Tab.~\ref{tab:params_description}.}
\label{fig:spins_plot}
\end{figure}

For a comprehensive visualization of the angles refer to Fig.~\ref{fig:ra_dec} and Fig.~\ref{fig:spins_plot}, and to Tab.~\ref{tab:params_description}.

\subsection{Luminosity distance}
The prior for the luminosity distance parameter varies according to different assumptions about the distribution of sources in the Universe. In the following, we will explore some common choices.
\subsubsection{\texttt{Power Law}}
In our comparison with LVK data, we used the \texttt{Power Law} prior for the luminosity distance, given by:
\begin{equation}
    \label{eq:uniform_d2}
    \pi(d_{L}) \propto d_L^2
\end{equation}

This choice implies that the number of GW sources is proportional to the volume, assuming a uniform distribution of sources in space without considering any redshift evolution. Since the volume of space scales with the square of the distance ($\mathrm{d}V = 4\pi d_L^2 \mathrm{d} d_L$), the probability density function is proportional to the square of the source distance. The distance we consider here is the distance we measure, i.e. the luminosity distance. 

Choosing this prior assumes that farther events are more likely simply because they cover a larger volume, without accounting for the expansion of the Universe.

\subsubsection{\texttt{Uniform in Comoving Volume}}
If we want to account for the expansion of the Universe, then we should choose the uniform in comoving volume prior:
\begin{equation}
    \label{eq:uniform_in_comoving_volume}
    \pi(d_L) \propto \frac{\mathrm{d}V_c}{\mathrm{d}d_L} = \frac{\mathrm{d}V_c}{\mathrm{d}z}\frac{\mathrm{d}z}{\mathrm{d} d_L}
\end{equation}
where we have made explicit the redshift dependence for easier calculations (see also Eq.~\ref{eq:uniform_in_source_frame}). 
For the derivative redshift with respect to luminosity distance we compute the inverse, which is easier:
\begin{equation}
    \frac{\mathrm{d}d_L}{\mathrm{d}z} = \frac{d_L}{1+z} + \frac{c}{H_0}\frac{1+z}{E(z)}
\end{equation}
where 
\begin{equation} 
    d_L = \frac{c}{H_0}(1+z)\int_0^z \frac{\mathrm{d}z'}{E(z')}
\end{equation}
and
\begin{equation} 
    E(z)=\sqrt{(1+z)^3\Omega_{\rm m} + \Omega_{\Lambda}}
\end{equation}
with $\Omega_{\rm m}$ and $\Omega_{\Lambda}$ being the dimensionless matter and dark energy densities in the $\Lambda$CDM framework.

Eq.~\ref{eq:uniform_in_comoving_volume} takes into account the geometry and the expansion of the Universe.
A \texttt{Uniform in Comoving Volume} prior favors redshifts where larger volumes are available, in accordance with the cosmological principle of a homogeneous and isotropic Universe.

\subsubsection{\texttt{Uniform in Source-Frame Time}}
Furthermore, if we want to assume a constant event rate per unit of time in the source frame, then we have to adjust for the shifting of source frame time:
\begin{equation}
    \label{eq:uniform_in_source_frame}
    \pi(d_{L}) \propto \frac{1}{1+z}\frac{\mathrm{d}V_c}{\mathrm{d}d_L} = \frac{1}{1+z}\frac{\mathrm{d}V_c}{\mathrm{d}z}\frac{\mathrm{d}z}{\mathrm{d}d_L}
\end{equation}
The $(1+z)$ factor accounts for the time dilation when passing from source to detector frame:
\begin{equation}
    \frac{\mathrm{d}t_{\rm source}}{\mathrm{d}t_{\rm detector}}=\frac{1}{1+z} 
\end{equation}

This prior ensures a uniform event rate in the source frame.

\begin{table*}[h!]
\caption{List of all the parameters used in our analysis. For a better understanding of some definitions, refer to Fig.~\ref{fig:ra_dec} and Fig.~\ref{fig:spins_plot}.}
\label{tab:params_description}
\begin{tabular}{l | c c}
\textbf{parameter}  &\textbf{label}  &\textbf{description}\\
\toprule
chirp mass   &${\mathcal{M}}_c$   &combination of detector-frame masses: ${\mathcal{M}}_c=\frac{(m_1m_2)^{3/5}}{(m_1+m_2)^{1/5}}$ 
\\
mass ratio  &$q$  &ratio between the secondary and the primary component masses, $0<q\le 1$ 
\\
luminosity distance &$d_L$ &luminosity distance to the source
\\
theta\_JN &$\theta_{JN}$ &angle between the line of sight $\vec{N}$ the total angular momentum of the binary $\vec{J}$
\\
declination &\texttt{DEC} &coordinate on celestial sphere corresponding to latitude on Earth
\\
right ascension &\texttt{RA} &coordinate on celestial sphere corresponding to longitude on Earth
\\
phase &$\phi$ &phase of the gravitational wave at the coalescence
\\
polarization angle &$\Psi$ &rotation of GW polarizations reference frame with respect to detector's arms in Earth's frame
\\
geocentric time &$t_c$ &merger time, calculated as GPS time
\\
a\_1 &$a_1$ &dimensionless spin magnitude of primary component
\\
a\_2 &$a_2$ &dimensionless spin magnitude of secondary component
\\
tilt\_1 &\texttt{tilt}$_{1}$ &angle between the spin vector of primary component and the orbital angular momentum $\vec{L}$
\\
tilt\_2 &\texttt{tilt}$_{2}$ &angle between the spin vector of secondary component and the orbital angular momentum $\vec{L}$
\\
phi\_12 &\texttt{phi}$_{12}$ &angle between the projections of the two spin vectors onto the orbital plane
\\
phi\_JL &\texttt{phi}$_{JL}$ &azimuthal angle of the orbital angular momentum $\vec{L}$ around the total angular momentum $\vec{J}$
\\
\bottomrule
\end{tabular}
\end{table*}

\bibliography{main}

\end{document}